# Room-temperature bulk plasticity and tunable dislocation densities in KTaO$_3$

Xufei Fang[1a*#], Jiawen Zhang[2#], Alexander Frisch[1a], Oliver Preuß[3], Chukwudalu Okafor[1a], Martin Setvin[4], Wenjun Lu[2*]

[1]Institute for Applied Materials, Karlsruhe Institute of Technology, 76131 Karlsruhe, Germany

[2]Department of Mechanical and Energy Engineering, Southern University of Science and Technology, 518055 Shenzhen, China

[3]Department of Materials and Earth Sciences, Technical University of Darmstadt, 64287 Darmstadt, Germany

[4]Departement of Surface and Plasma Physics, Charles University, 180 00 Praha, Czech Republic

*Corresponding authors: xufei.fang@kit.edu (X.F.); luwj@sustech.edu.cn (L.W.)

[#] These authors contribute equally to this work.

[a] Previously in Department of Materials and Earth Sciences at Technical University of Darmstadt where this work was initiated.

**Abstract**

We report room-temperature bulk plasticity mediated by dislocations in single-crystal cubic KTaO$_3$, contrasting the conventional knowledge that single-crystal KTaO$_3$ is susceptible to brittle cleavage. A mechanics-based combinatorial experimental approach using cyclic Brinell indentation, scratching, and uniaxial bulk compression consistently demonstrates room-temperature dislocation plasticity in KTaO$_3$ from the mesoscale to the macroscale. This approach also delivers tunable dislocation densities and plastic zone sizes. Scanning transmission electron microscopy analysis underpins the activated slip system to be <110>{1-10}. Given the growing significance of KTaO$_3$ as an emerging electronic oxide and the increasing interest in dislocations for tuning physical properties of oxides, our findings are expected to trigger synergistic research interest in KTaO$_3$ with dislocations.

## 1. Introduction

Ceramics are generally known for their brittleness at room temperature, primarily due to the lack of dislocation-mediated plasticity. However, recent advancements have demonstrated room-temperature dislocation plasticity in various ceramic materials at small scales using techniques like nanoindentation [1-3] and nano-/micropillar compression [4, 5]. These methods minimize flaw populations and favor plastic flow over cracking or suppress crack propagation by means of locally high compressive hydrostatic stress (as seen in nanoindentation with shallow depth) or by reducing deformed volumes (in nano-/micropillar compression).

In contrast, ceramics that exhibit bulk and mesoscale plasticity under ambient conditions are relatively rare, despite their longer research history. Crystals with rock-salt structures, such as LiF, NaCl, and KCl, have been extensively studied for their dislocation mechanics since the 1950s. Classic studies by Johnston and Gilman on LiF single crystals have systematically explored dislocation multiplication, nucleation, and mobility through dislocation etch pit studies [6-8]. Similarly, NaCl crystals have been pivotal in understanding dislocation-based fracture toughness [9, 10] and electro-plasticity as well as the charge of dislocations in ionic crystals [11]. Another notable group of ductile ceramics includes simple oxides with rock-salt structures, particularly single-crystal MgO. Discovered to be plastically deformable in bulk compression [12, 13], as early as in the late 1950s, MgO continues to be studied for its fundamental role in the Earth's lower mantle [14] and as a model system for understanding the elementary dislocation mechanics in oxides [15].

Due to the wide bandgap of the aforementioned ductile crystals, their application in electronic devices is limited. Consequently, more attention has been directed towards other ductile semiconductors such as ZnS [16] and perovskite oxides. In 2001, Brunner et al. [17] reported the *surprising* discovery of room-temperature plasticity in $SrTiO_3$ (cubic structure) perovskite oxide, demonstrating a plastic strain of up to ~7% under uniaxial bulk compression. $SrTiO_3$ was probably the first perovskite oxide discovered to exhibit bulk plasticity at room temperature. Owing to its prototypical nature in condensed matter physics [18, 19] and its role as a model electronic oxide, room-temperature dislocation plasticity in $SrTiO_3$ has thereafter been extensively studied, ranging from macroscale [20], mesoscale [21, 22], down to nanoscale [23-26]. Later in 2016, Mark et al. [27] reported *unexpected* room-temperature bulk plasticity in single-crystal $KNbO_3$, which is orthorhombic at room temperature [28]. These finding was further confirmed by Höfling et al. in 2021 [29] and Preuß et al. in 2023 [30]. So far, $SrTiO_3$ and $KNbO_3$ have remained the only two perovskite oxides reported in literature regarding room-temperature bulk plasticity.

In light of the rising interest in using dislocations as one-dimensional line defects in ceramic oxides to harvest both functional and mechanical properties [31, 32], there is a pressing need to seek more room-temperature ductile ceramics as well as to engineer dislocations into such functional oxides for harvesting dislocation-tuned properties. Recently, our team achieved this by developing a combinatorial experimental toolbox for tuning dislocation densities and plastic zone sizes [33] in e.g., single-crystal $SrTiO_3$. However, the pursuit in discovering more room-temperature ductile ceramics remains largely unexplored so far. With the abundant techniques and experimental protocols recently established for room-temperature dislocation engineering [33], we begin to ask if this toolbox can be used to discover other ceramics that can be plastically deformed at room temperature at meso-/macroscale.

In this work, we report the third ductile perovskite oxide (potassium tantalate, $KTaO_3$) for its room-temperature bulk plasticity and tunable dislocation densities and plastic zone size. $KTaO_3$ recently received much attention, owing to its potential for functional oxide electronics [34, 35] and tunable ferroelectricity that is achieved through Nb doping [36]. These physical properties extend applications of $KTaO_3$ towards piezocatalysis, pyrocatalysis [37], and photocatalysis [38, 39]. Dislocations represent an additional degree of freedom that translates into all these applications, and it is therefore likely to spark more research interest for dislocation-based functional properties in $KTaO_3$.

## 2. Experimental procedure

Synthetic single-crystal, undoped $KTaO_3$ samples were prepared by solidification from a nonstoichiometric melt in Oak Ridge National Laboratory. For cyclic Brinell indentation and scratching tests, a sample with a geometry of ~1 mm×5 mm×5 mm was used. The tests were run on a hardness indenter (Karl-Frank GmbH, Weinheim-Birkenau, Germany), following the experimental procedure established by the current authors [21, 22]. The indenter is mounted with a Brinell indenter with a diameter of 2.5 mm (hardened steel ball, Habu Hauck Prüftechnik GmbH, Hochdorf-Assenheim, Germany) and a movable stage (Physik Instrumente GmbH & Co. KG, Karlsruhe, Germany). For both cyclic indentation and scratching tests, a dead weight of 1 kg was used. For scratching tests, a speed of 0.5 mm/s (lateral motion) was adopted. Additionally, silicon oil was used as a lubricant to reduce the indenter wear and to suppress sample crack formation. After mechanical deformation, the sample was cleaned with acetone and dried in air. The surface slip traces in the plastic zones were visualized using a laser confocal microscope (LEXT OLS4000, Olympus IMS, Waltham, USA). Dark-field imaging mode was used on a Zeiss optical microscopy (Zeiss Axio Imager2, Carl Zeiss AG, Oberkochen, Germany) to exclude potential crack formation underneath the surface.

For uniaxial bulk compression, single-crystal $KTaO_3$ samples (Hefei Single Crystal Material Technology, Anhui, China) with a dimension of 3 mm×3 mm×6 mm were used. The long axis of the samples was aligned in the [001] direction and the other two surfaces being (100) and (010). The samples were six-side polished to mirror finish to minimize surface damage. Uniaxial compression was performed with a constant strain rate of $1.5 \times 10^{-4}$ s$^{-1}$ (MTS E45, USA). The deformation images were collected in the VIC-gauge 2D software (Correlated Solution, Inc., USA). Before the compression testing, two $Al_2O_3$ plates were placed on top and bottom of the $KTaO_3$ specimen to offer smoother contact surfaces.

To visualize the dislocation structures in the plastic zone, TEM (transmission electron microscopy) lamellae were lifted out along the scratching direction, using focused ion beam (FIB). The lamellae with a thickness of ~80 nm were prepared and thinned by a FIB (Helios Nanolab 600i, FEI, Hillsboro, USA). Microstructures of $KTaO_3$ including dislocations were characterized by a 200 kV-TEM (FEI Talos F200X G2, Thermo Fisher Scientific, USA) with STEM mode. The ABF-STEM (annular bright field-scanning transmission electron microscopy) images were collected with inner and outer semi-collection angles of 12-20 mrad.

## 3. Results & Analyses

### 3.1. Brinell indentation & scratching

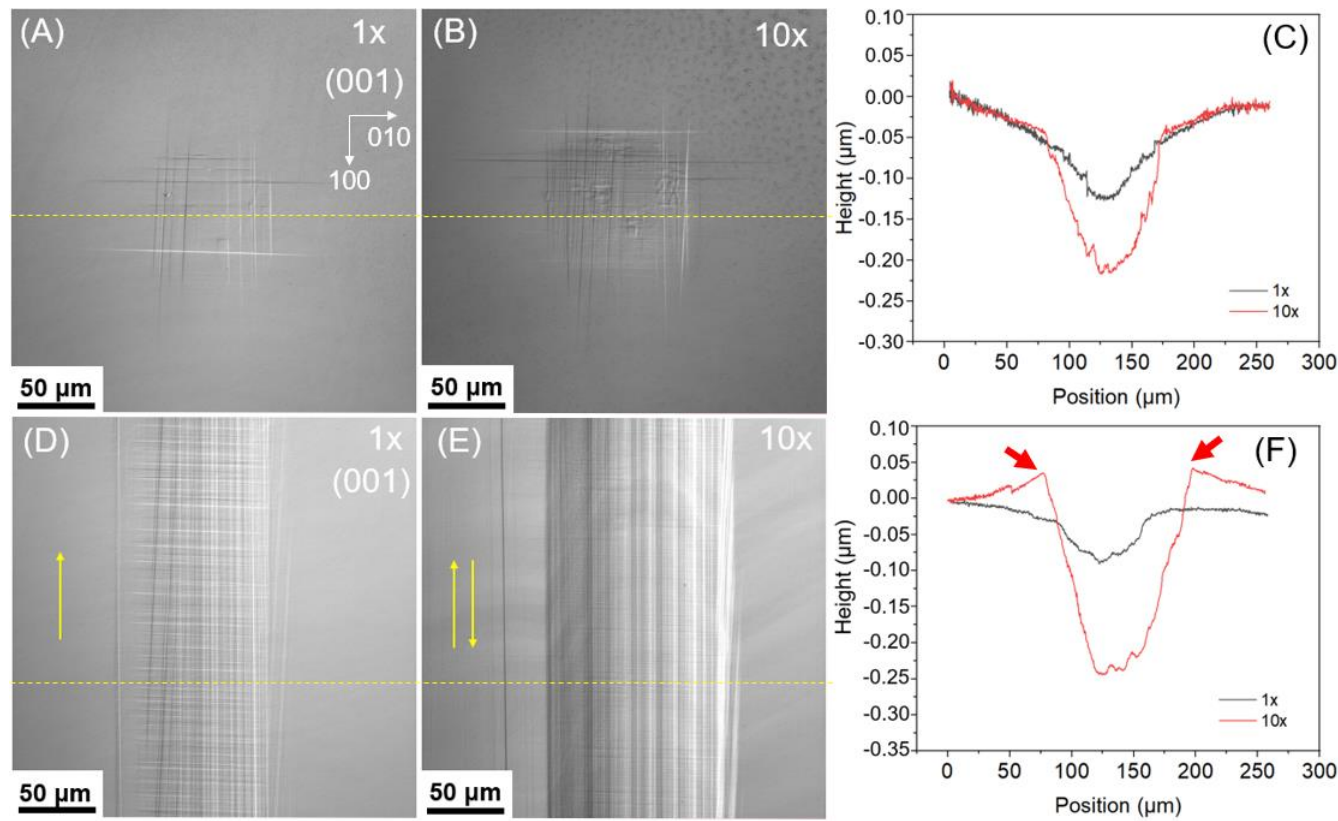

**Fig. 1** Laser microscopy images featuring the plastic deformation on the (001) surface after Brinell ball indentation (A-B) and scratching (D-E). Note 1x and 10x stand for 1-cycle and 10-cycle deformation, respectively. The yellow arrows in (D, E) indicate the scratching direction. Depth profiles (C, F) corresponding to the yellow dashed lines were extracted, with the unperturbed sample surface corresponding to the zero point in the y-axis.

**Figure 1A-C** demonstrate the surface plastic deformation after 1 cycle (1x) and 10 cycles (10x) Brinell indentation. Notice that the surface slip traces are aligned vertical and horizontal after 1x and 10x indentation (**Fig. 1A**), and the slip trace number increases from 1x to 10x indentation (**Fig. 1B**). The depth profiles of these two indentation imprints in **Fig. 1C** (corresponding to the two yellow dashed lines in **Fig. 1A, B**) indicate a maximum depth of ~150 nm after 1x and ~250 nm after 10x indentation. As the plastic zone size in both cases has a diameter of ~150 μm, this suggests that the surface is nominally flat in the indented region.

In addition to the cyclic Brinell indentation test, cyclic scratching tests were performed using the same indenter. **Figure 1D, E** reveal an increase in the slip trace densities with the increasing scratching number from 1x to 10x. The maximum depth of the scratch tracks (**Fig. 1F**, corresponding to the two yellow dashed lines in **Fig. 1D, E**) was measured to increase from ~150 nm (1x) up to ~600 nm (10x). Different from the sink-in feature in the 1x as well as the 1x and 10x indentation imprints, note that the 10x depth profile exhibits two shoulders (pile-up, indicated by the two red arrows in **Fig. 1F**). This pile-up was likely caused by the "plastic plowing" of the material by the spherical indenter during the back-and-forth cyclic scratching (scratching directions indicated by the yellow arrows in **Fig. 1E**). This pile-up is also strong evidence of good room-temperature plastic deformation of this material.

To rule out the possible cracking underneath the indentation imprints/scratch track, dark-field imaging mode (sensitive to under-surface cracks, featured as white contrasts) was used. No visible cracks were found up to 25x cycles of scratching. This observation is consistent with the results obtained on other ductile oxides (e.g., $SrTiO_3$ [21, 22]) at room temperature. Worth mentioning is that, both the Brinell ball indentation and scratching test results closely resemble the slip trace features as in single-crystal $SrTiO_3$ with the same (001) surface being deformed. The plastic zone size is also almost identical for $KTaO_3$ observed here as in $SrTiO_3$ [21] under the same loading conditions. Such similarities strongly suggest that the lattice friction stress in $KTaO_3$ shall be sufficiently low to allow easy dislocation glide and multiplication at room temperature, as in the case of $SrTiO_3$. Note that single-crystal $SrTiO_3$ was reported to plastically yield around 110-150 MPa during bulk compression along the [001] orientation (summarized literature results by Stich et al. [40]). If the deformation behavior is similar for $SrTiO_3$ and $KTaO_3$, then the yield strength of single-crystal $KTaO_3$ should be around the same range. To this end, further validation with uniaxial bulk compression is then performed on $KTaO_3$ along the [001] direction, as demonstrated in **Fig. 2**.

### 3.2. Uniaxial bulk compression

The engineering stress-strain curve in **Fig. 2A** as well as the deformation process in **Fig. 2B** of single-crystal $KTaO_3$ clearly demonstrate the bulk plastic deformation. Unlike the room-temperature bulk stress-strain curves in $SrTiO_3$ [17, 41] and $KNbO_3$ [27, 29], the stress-strain curve for $KTaO_3$ exhibits an upper yield point and a lower yield point (indicated in **Fig. 2A**). The upper yield point (**Fig. 2A**) was found to be $\sigma_{y\_up}$ = 274 MPa while the lower yield point $\sigma_{y\_low}$ = 260 MPa. This load drop behavior is not uncommon for ceramics that undergo plastic deformation. For instance, similar yield behavior was reported in bulk compression of sapphire at high temperature [42] and discussed in single-crystal LiF at room temperature [43]. Due to the initially low mobile dislocation density in such ceramic crystals, dislocation multiplication was proposed to have caused such a stress drop but not due to the unpinning of a Cottrell atmosphere as in the case of α-iron, where dislocation pinning results from the impurities such as carbon [44].

After the load drop followed by the upper yield point, the stress increased from ~250 MPa (i.e., the lower yield point in **Fig. 2A**) up to ~300 MPa, where another load drop was observed. On the one hand, the stress increase indicates work hardening behavior, as evidenced by the increased number of slip traces (dark lines that lie 45° to the loading axis, indicated by the black arrows) in **Fig. 2B3-B4**. These slip traces confirm that the dislocation glide occurs on the {110} planes. On the other hand, the load drop was found to correspond to the primary fracture event observed by the *in-situ* deformation (**Fig. 2B5,** where the bright features on the top right corner indicate a main crack formed, see red arrow). The deformed sample was captured (**Fig. 2B6**, at a strain of ~6.3%) right before the sample shattered into pieces at a maximum strain of ~6.5%. Note the sample in **Fig. 2B6** is already full of cracks as reflected by the white contrast.

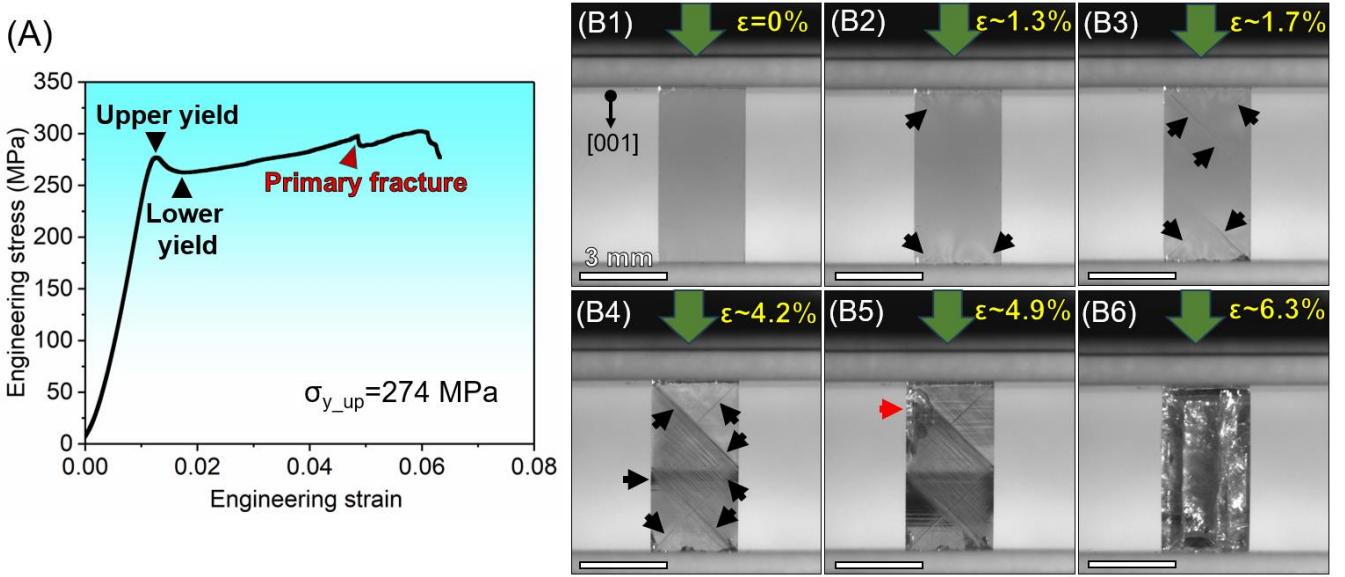

**Fig. 2** Bulk compression of single-crystal $KTaO_3$ along the [001] direction: (A) engineering stress-strain curve; (B1-B6) Screenshots of the *in-situ* bulk compression at different strains. The black arrows indicate the slip traces and the red arrow indicates the crack formation. The scale bar in (B1) is consistent for all six sub-figures.

The observed yield stress during bulk compression agrees with the stress analysis using Hertzian contact for the Brinell ball indentation in **Fig. 1**, as will be rationalized in the following: Consider that the post-mortem plastic zone imprint (**Fig. 1A, B**) has a diameter of $D$ = 150 µm, with the load of $P$ = 1 kg (9.8 N), we estimate the mean pressure $p_0$ underneath the indenter to be ~555 MPa using the expression $p_0 = P/(\pi D^2/4)$ [45]. The maximum shear stress (upon plastic yield) during spherical indentation can be calculated according to Swain & Lawn [45], yielding a value of about $\tau_{max}$ = 255 MPa ($0.46p_0$). Note that the critical resolved shear stress ($\tau_{CRSS}$) in the current uniaxial bulk compression is half of the value of the upper yield strength ($\sigma_{y\_up}$ = 274 MPa in **Fig. 2A**), giving $\tau_{CRSS}$ = 137 MPa. Consider that the estimation above is made on the Brinell indentation imprint, which is already in the plastic deformation regime while Hertzian contact theory is for elastic deformation, the $\tau_{max}$ is expected to be larger than $\tau_{CRSS}$. As both the $\tau_{max}$ and $\tau_{CRSS}$ are much smaller than the theoretical shear strength (~19 GPa for $KTaO_3$, estimated by $G/2\pi$, $G$ is the shear modulus) for homogeneous dislocation nucleation, this suggests that the plastic deformation under the large Brinell indenter as well as the uniaxial bulk compression is mediated by dislocation multiplication and dislocation glide. It is thus expected that the lattice friction stress for dislocation glide in KTaO$_3$ at room temperature shall be close to $\tau_{CRSS}$ = 137 MPa (likely the upper bound), which is higher but close to that in SrTiO$_3$ (~90 MPa [46]) and KNbO$_3$ (~30 MPa [27, 29]).

**3.3. TEM characterization of dislocations**

To directly prove that the dislocation densities in the plastic zone can be tuned by increasing the number in the deformation cycles, we performed TEM analysis in the 1x and 10x scratched regions. As illustrated in **Fig. 3A**, there are few dislocations with long segments in the area of view after 1x scratching. These long segments are aligned on the {110} planes. This is consistent with the slip trace observation during bulk deformation (**Fig. 2B**). After 10x scratching, the dislocation density increased dramatically and the dislocation lines are heavily tangled up with each other, where the dislocation density is estimated to be higher than 10$^{14}$/m$^2$. Although the majority of the dislocations are still projected in the [110] directions, there are many short and curved dislocation segments generated. These new features are clearly a result of the cyclic scratching, which strongly promotes the dislocation multiplication and interaction. Such profuse dislocation multiplication significantly promotes the dislocation plasticity, which is in line with the observation of the pile-up behavior after 10x Brinell ball scratching. Further analysis of the Burgers vector and the line vector of the dislocations suggest a Burgers vector of <110> and both edge and screw types of dislocations are generated. It is confirmed that KTaO$_3$ has the same slip system as in the case of SrTiO$_3$ [18] and KNbO$_3$ (when considered pseudo-cubic) [27], namely, the <110>{1-10} slip systems are activated at room temperature.

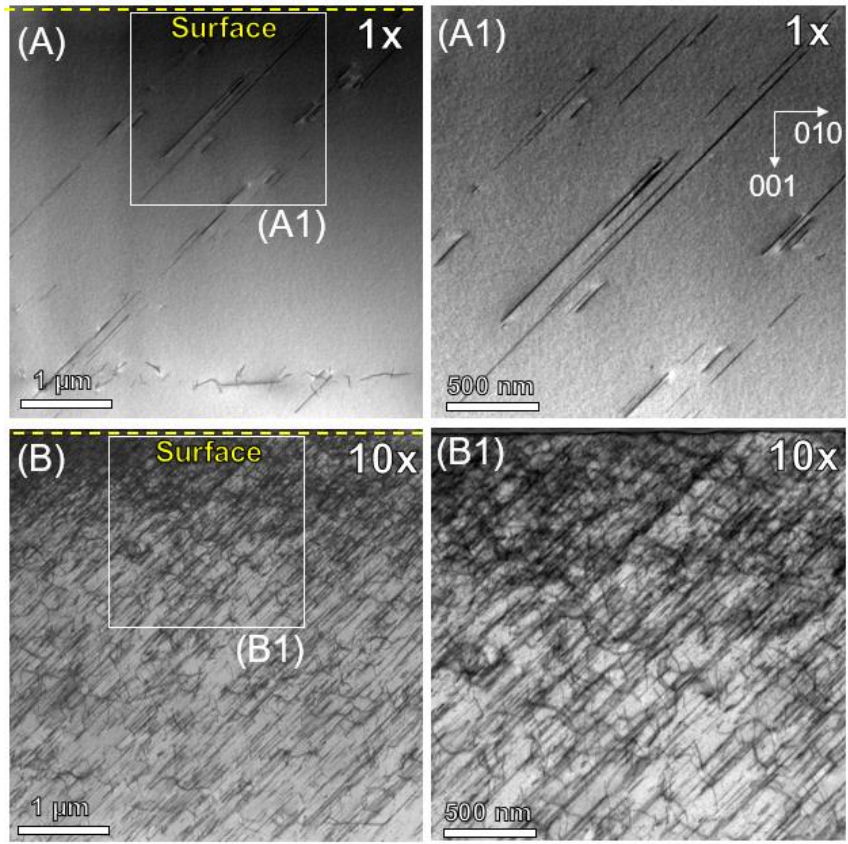

**Fig. 3** Visualization of dislocations (dark lines) in ABF-STEM (annular bright field-scanning transmission electron microscopy): (A) after 1x scratching; (B) after 10x scratching. The dislocation density is tunable depending on the number of cyclic scratching. The orientations in (A1) is consistent for all four sub-figures.

## 4. Discussion

Single-crystal KTaO$_3$ was reported to be very susceptible to brittle cleavage along the (001) surface [47]. The results here may seem counterintuitive as we are successful in engineering dislocations in both mesoscale (Brinell ball indentation/scratching) and bulk compression without inducing visible cracks. The choice of KTaO$_3$ was made based on its structural and atomic similarities to other two ductile perovskite oxides SrTiO$_3$ and KNbO$_3$. KTaO$_3$ has a cubic structure at room temperature as in SrTiO$_3$, also Ta and Nb have the same ionic radii 0.64 Å (for +5 charge state and 6-fold coordination) [48]. This leads to the hypothesis that KTaO$_3$ might also be ductile at room temperature as SrTiO$_3$ and KNbO$_3$.

At this stage, the underlying mechanisms for yielding such dislocation plasticity in single-crystal KaTiO$_3$ remain unclear. Nevertheless, considering that both KTaO$_3$ and SrTiO$_3$ are very similar in crystal structure and much often used together for comparison in their physical properties [34], it is likely that the dislocation mechanisms in both materials may be analogous, particularly concerning the dislocation core structure. For instance, TEM observations and atomistic simulations in SrTiO$_3$ [49, 50] as well as in KNbO$_3$ [28] suggest that the dislocations are dissociated into partials, which facilies good dislocation mobility at room temperature. To confirm if this would be the same for KTaO$_3$, future work will involve detailed TEM characterization as well as molecular dynamics simulations in this direction. It is worth noting that, due to the earlier discovery of room-temperature bulk dislocation plasticity and the simple dislocation introduction process in SrTiO$_3$, most of the dislocation-based functional and mechanical properties studies [51-53] have been reported using this model system. Now with this simple and efficient dislocation engineering demonstrated here in KTaO$_3$, and considering that KTaO$_3$ has been deemed as the "new kid on the spintronics block" [34], it is expected that our finding will serve as a fundamental building block for upcoming versatile studies in KTaO$_3$ as in the case of SrTiO$_3$ discussed above.

## 5. Conclusion

We found single-crystal KTaO$_3$ with a cubic structure can be plastically deformed at room temperature via bulk uniaxial compression, Brinell ball indentation and scratching. The room-temperature slip systems in KTaO$_3$ are identified to be <110>{1-10}, the same as those observed in SrTiO$_3$ and KNbO$_3$ deformed at room temperature. For the single crystals compressed along the <001> direction in this work, the bulk yield strength of KTaO$_3$ was ~274 MPa, and the critical resolved shear stress is estimated to be ~137 MPa to move the dislocations. Unlike the discrete slip bands generated during bulk compression, the Brinell ball indentation and scratching generate continuous plastic zones extending up to hundreds of micrometers with dislocation densities exceeding ~10$^{14}$/m$^2$ after 10-cycle scratching. It remains an open question at this stage to pinpoint the fundamental mechanisms responsible for the room-temperature dislocation plasticity in KTaO$_3$. Our findings are expected to open new avenues to investigate dislocation-tuned mechanical and functional properties in KTaO$_3$.

**Acknowledgment**

X. Fang and A. Frisch acknowledge the European Research Council (ERC) under Grant No. 101076167 (MECERDIS) for supporting the research. Views and opinions expressed are, however, those of the authors only and do not necessarily reflect those of the European Union or the European Research Council. Neither the European Union nor the granting authority can be held responsible for them. C. Okafor acknowledges the financial support by the Deutsche Forschungsgemeinschaft (DFG, grant No. 510801687). O. Preuß thanks the DFG for the funding (grant No. 414179371). M. Setvin acknowledges the support from the Czech science foundation, project GACR 20-21727X. W. Lu is supported by Shenzhen Science and Technology Program (grant no. JCYJ20230807093416034), National Natural Science Foundation of China (grant no. 52371110) and Guangdong Basic and Applied Basic Research Foundation (grant no. 2023A1515011510). The authors acknowledge the use of the facilities at the Southern University of Science and Technology Core Research Facility. M. Setvin and X. Fang would like to thank L. A. Boatner at Oak Ridge National Laboratory for providing the KTaO$_3$ crystal for the indentation tests. We thank Prof. Rödel at TU Darmstadt for the discussion.

**Authors contribution:**

X.F.: conceptualization, experimental design, samples acquisition, indentation and scratching tests, writing first draft; W. L., J. W.: experimental design, bulk deformation, and TEM analysis; C.O., O. P.: Imaging and surface characterization; A.F.: sample preparation, indentation and scratching tests; M.S.: sample preparation, discussion. All authors contributed to the revision of the draft.